\documentclass[utf8]{FrontiersinHarvard} % for 
\usepackage{url,hyperref,lineno,microtype,subcaption}
\usepackage[onehalfspacing]{setspace}

\usepackage{multirow}
\usepackage{graphicx}
\usepackage{subcaption}
\usepackage{mathptmx}       % Times Roman
\usepackage{helvet}         % Helvetica
\usepackage{txfonts}
\usepackage{caption}
\usepackage{array}
\usepackage{makecell} 
\usepackage[margin=1in]{geometry}

\def\keyFont{\fontsize{8}{11}\helveticabold }
\def\firstAuthorLast{Ghatul {et~al.}} %use et al only if is more than 1 author
\def\Authors{{Shubham Ghatul*$^{1,4}$, Rekhesh Mohan$^1$, Jayant Murthy$^1$, Margarita Safonova$^{1,2,3}$, Praveen Kumar$^{1,5}$, Maheswar Gopinathan$^1$, Shubhangi Jain$^{1,4}$, Mahesh Babu S.$^1$}
}

\def\Address{
$^{1}$Indian Institute of Astrophysics, Bangalore, India,\\
$^{2}$M.P. Birla Institute of Fundamental Research, Bangalore, India\\
$^{3}$LUNEX EuroMoonMars EuroSpaceHub, Leiden,  Noordwijk, The Netherlands\\
$^{4}$Department of Applied Optics and Photonics, University of Calcutta, Kolkata, India\\
$^{5}$Indian Institute of Science, Bangalore, India}
\def\corrAuthor{Shhubham Ghatul}
\def\corrEmail{shubham.jankiram@iiap.res.in}

\begin{document}
\onecolumn
\firstpage{1}

\title[NUTEx]{{Near Ultraviolet Transient Explorer (NUTEx): A CubeSat-Based NUV Imaging Payload for Transient Sky Surveys}} 

\author[\firstAuthorLast ]{\Authors} %This field will be automatically populated
\address{} %This field will be automatically populated
\correspondance{} %This field will be automatically populated

\extraAuth{}% 
\maketitle

\begin{abstract}
The Near Ultraviolet Transient Explorer (NUTEx) is a CubeSat-based near-ultraviolet (NUV) imaging payload designed for transient sky surveys and is currently under development. CubeSats are compact and cost-effective satellite platforms that have emerged as versatile tools for scientific exploration and technology demonstrations in space. NUTEx is an imaging telescope operating in the 200–300 nm wavelength range, intended for deployment on a micro-satellite bus. The optical system is based on a Ritchey–Chrétien (RC) telescope configuration, featuring a 146-mm primary mirror. The detector is a photon-counting microchannel plate (MCP) device with a solar-blind photocathode, paired with an in-house-developed readout unit. The instrument has a wide field of view (FoV) of 4°, a peak effective area of approximately 18 cm² at 260 nm, and can reach a sensitivity of 21 AB magnitude (SNR = 5) in a 1200-second exposure. The primary scientific objective of NUTEx is to monitor the night sky for transient phenomena, such as supernova remnants, flaring M-dwarf stars, and other short-timescale events. The payload is currently scheduled for launch in Q2-2026. This paper presents the NUTEx instrument design, outlines its scientific goals and capabilities, and provides an overview of the electronics and mechanical subsystems, including structural analysis.

\tiny
 \keyFont{ \section{Keywords: CubeSats, Small Payloads, UV Instrumentation, Imaging telescope, Payload structure, mechanical design} }
\end{abstract}

\section{Introduction}

\noindent CubeSats are standardized miniature satellites that have emerged as a transformative platform for space-based astrophysical research, particularly in observational regimes such as the ultraviolet that are inaccessible from the ground. Their ability to deliver rapid, flexible, and cost-effective access to space allows them to complement larger observatories and respond swiftly to transient events. The growing availability of commercial-off-the-shelf (COTS) components, ranging from power systems to precision attitude control units, has further streamlined CubeSat development, significantly reducing both cost and time \citep{Douglas2019-rk, Spence2022-ad}. With mission budgets typically in the range of US\$5–10 million and development cycles of just 2–3 years, CubeSats enable rapid mission turnaround, opening new opportunities for focused science, technology demonstrations, and broader participation in space science by smaller institutions and emerging space programs \citep{Shkolnik2018-uv,WOELLERT2011663}.

We, the Space Payload Group (SPG) at the Indian Institute of Astrophysics (IIA), are dedicated to designing and building innovative, low-cost instruments for ultraviolet (UV) astronomy. Our designs and results have been published in the open literature and adopted by other groups. For example, our recent payloads \citep{SING_ExA} showcase our expertise in compact space instruments, including a star camera built around a Raspberry Pi \citep{starberry2}. These efforts illustrate that compact, resource-efficient instruments can still address important scientific questions in UV astronomy.

The Near Ultraviolet Transient Explorer (NUTEx) is a near-ultraviolet (NUV) imaging telescope currently under development to fly on a microsatellite bus. The instrument is a Ritchey–Chrétien (RC) telescope with a 146-mm diameter aperture, covering the 200–300 nm wavelength band using a solar-blind photocathode. It has a mean effective area of 15 cm\textsuperscript{2} across this band and can detect objects as faint as 21 AB magnitude (SNR = 5) with a 1200-second exposure. With a wide field of view (4°), NUTEx is optimized to observe and follow up variable NUV events such as stellar flares, supernova explosions, and other transient astrophysical events. NUTEx is particularly well suited to observe bright sky regions, like the Galactic plane, that are typically avoided by larger missions due to the sensitivity of their detectors. This enables the instrument to capture data from underexplored regions of the UV sky.

\section{Motivation and scientific goals}

\noindent The ultraviolet sky, particularly in the time domain, remains relatively underexplored despite its potential to provide important insights into explosive and variable astrophysical phenomena. Earlier missions provided valuable data but were constrained by narrow fields of view and a focus on targeted observations. Their sensitivity limits also prevented observations of brighter regions of the sky \citep{GALEX_BrightLimit1, GALEX_BrightLimit2}, such as the Galactic plane. These factors have left gaps in the systematic monitoring of transients and variables in the NUV, thereby limiting their ability to conduct broad time-domain surveys. This highlights the need for a dedicated UV survey instrument with wide-field capabilities. NUTEx, with its relatively lower sensitivity, is well suited to observe and survey these brighter regions of the sky that have been inaccessible to previous instruments.

Understanding the final stages of massive stars lives, particularly the processes leading to supernova (SN) explosions, remains an active area of research. While direct detections of SN progenitors are possible in rare, nearby cases, early UV observations provide an alternative way to probe stellar properties such as radius and surface composition before explosion \citep{Chevalier1992}. These early signals, including shock breakout and the subsequent UV shock-cooling emission, evolve rapidly, especially in compact stars, and are therefore inaccessible to most ground-based surveys \citep{Nakar2010}. A wide-field NUV imager can detect such fast transients, including breakout flares from extended stars and those embedded in dense circumstellar material \citep{ofek2010}, and follow their shock-cooling signatures to help constrain parameters such as explosion energy, ejecta mass, and extinction \citep{rabinak2011}. With broad sky coverage, such an instrument would enable observations of a larger sample of these events, contributing to a better understanding of the final stages of massive stellar evolution \citep{Sagiv2014}.

These objectives, focused on surveying and mapping the sky for transient events in the NUV, define the primary science goals of NUTEx. The instrument’s wide field of view and tolerance to bright regions were chosen to support these goals, which remain the central driver of the mission.

Another key scientific objective that NUTEx can address is the study of stellar flaring events, including those occurring on M-dwarf stars. Stellar flares are transient bursts of radiation caused by magnetic reconnection in stars with convective envelopes, releasing energy across nearly all wavelengths \citep{M_dwarf1}. Studying these flares in the NUV is important to capture emission components that are underestimated in optical bands, which allows more accurate energy assessments and a better understanding of their impact on stellar and close-in exoplanet environments \citep{Brasseur2019}. The Vera C. Rubin Observatory’s LSST is expected to detect large numbers of such flares in the optical domain, enabling statistical studies of their frequency, duration, and spatial distribution \citep{M_dwarf2}. However, the NUV emission, which arises from the hotter layers of the stellar atmosphere during flaring events, remains largely unexplored. NUTEx, with its wide field of view and tolerance to bright regions, may provide the missing higher energy coverage of these flare signatures and complement LSST by extending flare studies into the NUV region. Together, this combined approach may improve estimates of flare energetics and activity patterns in M-dwarf stars.

Although NUTEx is primarily designed for wide-field NUV surveys to detect transients and variable sources across large regions of the sky, it is also capable of responding to rapid alerts from rare events, such as GW170817 \citep{Evans_2017}. This event was a luminous and rapidly fading UV transient associated with a binary neutron star merger, with emission peaking around the wavelength range where NUTEx has maximum effective area, making the instrument suitable for such observations. Rapid follow-ups by NUTEx can complement facilities specifically dedicated to gravitational wave follow-up, such as QUVIK\citep{QUVIK}, by providing additional sky coverage and timely early UV measurements.

Feasibility of detecting near-Earth objects (NEOs) in the NUV band has been demonstrated in earlier studies \citep{Sagiv2014}, and wide-field surveys such as the Palomar Transient Factory (PTF) have shown the ability to discover small, fast-moving objects using real-time pipelines \citep{PTFS1, PTFS2}. While NEO detection is not a primary science goal for NUTEx, these studies highlight how incidental discoveries of such objects can naturally arise in wide-field survey observations. In the NUV, however, such capability is complicated by the sudden appearance of fast or bright transients in the field of view, which has previously triggered Bright Object Protection systems (e.g., in HST/STIS \citep{STIS_BOD} and AstroSat/UVIT \citep{UVIT_BOD1, UVIT_BOD2}). Some of these triggers have been attributed to unexpected non-celestial objects, possibly NEOs.

In contrast, NUTEx is designed with a wider FoV and improved tolerance to bright regions, enabling it to handle such events without detector saturation. This characteristic is presented here not as a dedicated science driver, but as a demonstration of the instrument’s robustness: NUTEx can continue uninterrupted UV observations even when confronted with bright or fast-moving objects that might otherwise compromise detector safety. This resilience directly supports its main science objectives by ensuring long-duration, wide-field monitoring of transients and variable sources without unplanned interruptions. A comparison of NUTEx with relevant existing and proposed missions is provided in Table\ref{tab:mission_comparison}, highlighting differences in wavelength coverage, survey strategy, and other mission parameters. Mission parameters for other facilities, such as cost and mass, are taken from recent official releases and may be subject to change as these missions proceed through development and launch phases.

\begin{table}[ht]
\renewcommand{\arraystretch}{1.5} % Row height
\setlength{\tabcolsep}{6pt} % Column spacing
\centering
\caption{Comparison of key mission parameters between NUTEx and other UV missions.}
\label{tab:mission_comparison}
\begin{tabular}{|p{2.8cm}|p{2.8cm}|p{2.8cm}|p{2.8cm}|p{2.8cm}|}
\hline
\textbf{Mission Parameters} & \textbf{ULTRASAT} & \textbf{QUVIK (NUV)} & \textbf{UVEX (NUV)} & \textbf{NUTEx} \\
\hline
\textbf{Type} & Wide-field UV survey & Rapid ToO (GW follow-up) & Deep UV follow-up & Wide-field bright-sky survey \\
\hline
\textbf{Wavelength range (nm)} & 230--290 & 260--360 & 220--270 & 200--300 \\
\hline
\textbf{Aperture (mm)} & 330 & 330 & $\sim$750 & 146 \\
\hline
\textbf{Detector} & 4$\times$(4K$\times$4K) CCD & (4K$\times$4K) CMOS & CMOS & MCP + CMOS \\
\hline
\textbf{Spatial resolution} & 8.3$''$ & $\sim$2.5$''$ & $\sim$2.25$''$ & $\sim$13$''$ \\
\hline
\textbf{Field of View (deg$^2$)} & $\sim$200 & $\sim$3.15 & $\sim$12 & ~12 \\
\hline
\textbf{Weight (kg)} & $\sim$160 & $\sim$130 & $\sim$300 & $\sim$ ~4.3 \\
\hline
\textbf{Approx. Cost (USD)} & $\sim$90M & $\sim$10M & $>$300M & $\sim$0.5M \\
\hline
\textbf{Current Status} & Under development (launch $\sim$2027) & Under development (launch 2028-2029) & Proposed & Ready for launch by mid-2026 \\
\hline
\end{tabular}
\end{table}

\section{The instrument}

\noindent The optical design of NUTEx has been developed to meet its primary science goal of surveying the transient sky in the NUV with a wide field of view capability. For the structure of the instrument, we have nominally chosen microsatellite platforms provided by the Indian Space Research Organization (ISRO), which constrain the instrument volume to approximately 12U and impose a no-moving-parts requirement. The combination of these constraints and the intrinsically low UV photon flux from our targets required an optical design that minimizes reflective losses and maximizes throughput, with efficiency in the NUV band guiding the choice of coatings and materials. The mirrors are made from Zerodur, the lenses from fused silica, and the fixtures from aluminum alloy Al 6061-T6, providing stability while staying within the allocated mass and power budgets.

\begin{figure}
    \centering
\includegraphics[width=0.65\linewidth]{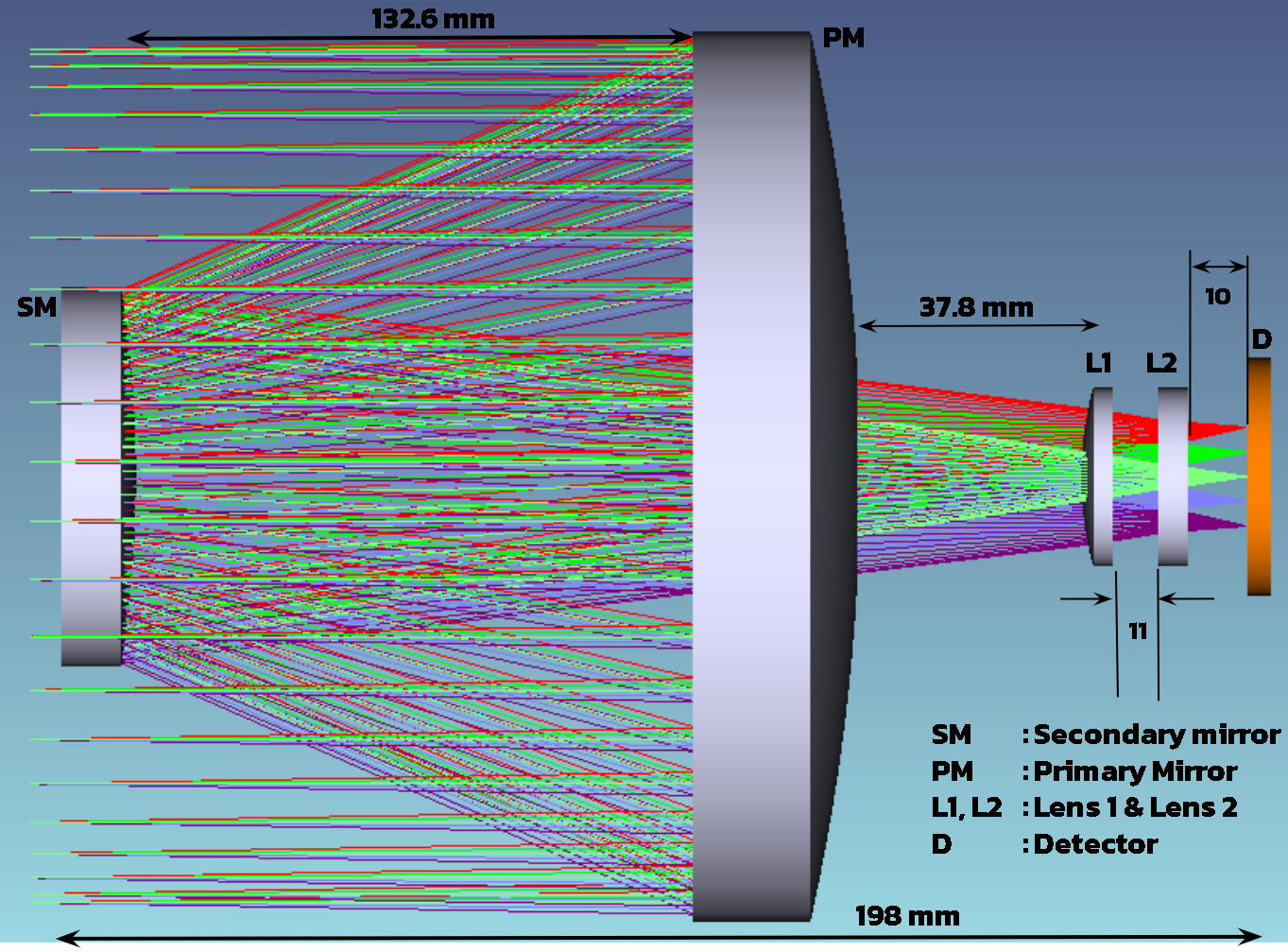}
    \caption{Optical design of NUTEx instrument.}
    \label{fig:OD_NUTEx}
\end{figure}

\subsection{Optical design}

\noindent To minimize reflection losses, the optical system uses only two reflecting surfaces: the primary mirror and the secondary mirror. The wide-field Ritchey-Chrétien configuration introduces additional optical aberrations, which are corrected by a set of two identical lenses positioned before the focal plane. An MCP-based photon-counting detector (PCD) is placed at the focal plane of the telescope. The optical design has been developed in coordination with the structural constraints and detector specifications, and the following sections provide a detailed discussion of the key components.

\begin{table}[ht]
\renewcommand{\arraystretch}{1.4} % Row height
\setlength{\tabcolsep}{10pt} % Column spacing
\centering
\caption{Key specifications of NUTEx  optical components}
\label{tab:optical_specs}
\begin{tabular}{|c|c|c|c|}
\hline
\textbf{Parameter} & \textbf{Primary Mirror} & \textbf{Secondary Mirror} & \textbf{Lens System} \\
\hline
\textbf{Type} & On-axis hyperbolic & On-axis hyperbolic & Corrector lenses \\
\hline
\textbf{Diameter} & 146 mm & 64 mm & 30 mm  \\
\hline
\textbf{Thickness} & 28 mm & 7.8 mm & 5 mm \\
\hline
\textbf{Central obscuration} & 34 mm dia. & NA & NA \\
\hline
\textbf{Conic constant} & $-1.237 \pm 0.005$ & $-6.56 \pm 0.005$ & - \\
\hline
\textbf{Radius of curvature} & $-335.28 \pm 0.1$ mm & $-201.23 \pm 0.5$ mm & $-60.28 \pm 0.5$ mm \\
\hline
\textbf{Substrate} & Zerodur & Zerodur & Fused silica \\
\hline
\textbf{Coating} & Al + MgF\(_2\) & Al + MgF\(_2\) & MgF\(_2\) \\
\hline
\textbf{Surface roughness} & $\lambda/10$ & $\lambda/10$ & $\lambda/10$ \\
\hline
\textbf{Mass} & 1035 gm & 150 gm & 50 gm (2 Nos) \\
\hline
\end{tabular}
\end{table}

The primary mirror (PM) is a 146-mm diameter on-axis hyperbolic mirror, fabricated from Zerodur to minimize thermal expansion and ensure stability in space environment. It is coated with aluminum and overlaid with an MgF\(_2\) layer for enhanced durability and reflectivity in the UV. The secondary mirror (SM) is also on-axis hyperbolic mirror, matched to the primary for optical compatibility. Aberrations introduced by the wide-field design are corrected using two identical fused silica lenses placed before the focal plane. The specifications of these optical components are summarized in Table~\ref{tab:optical_specs}.

The detector system consists of a 40-mm MCP-based PCD developed by Photek\footnote{\url{http://www.photek.com/}}, paired with a Raspberry Pi (RPi) based readout unit. This compact readout architecture, as demonstrated by \citet{readout}, successfully integrated and qualified for the SING payload~\citep{SING_P}, is well-suited for CubeSat-class instruments. Together, the detector and the readout form a complete photon-counting imaging assembly for NUTEx.

\subsubsection{Design analysis}

The field of view of ~4° in NUTEx arises directly from the optical and detector configuration. With an effective focal length of 476 mm (f/3.3), a 146 mm primary mirror, a 64 mm secondary mirror, and a 40 mm diameter circular detector at the focal plane, the resulting coverage is limited to ~4°. Expanding the FoV would either require shortening the focal length, which would degrade image quality and make aberration control more difficult in a Ritchey Chrétien configuration, or increasing the detector size, which is constrained by practical considerations. The adopted parameters therefore represent a balance between optical feasibility, detector coverage, and aberration control.

\begin{figure}[htbp]
\begin{minipage}[t]{0.4\linewidth}
\centering
\includegraphics[width=\linewidth]{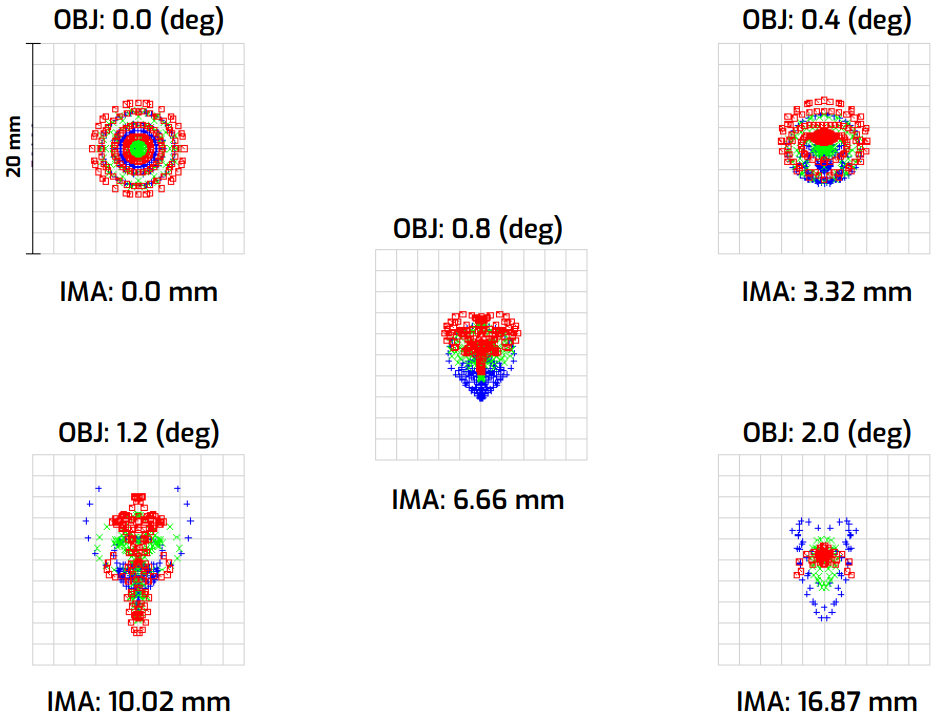}
\caption{Spot diagrams at the telescope focal plane for different field angles (0\textdegree~to 2\textdegree) marked with different wavelengths (blue: 2000~\AA, green: 2500~\AA, red: 3000~\AA).}
\label{fig:spot diagram}
\end{minipage}
\hfill
\begin{minipage}[t]{0.55\linewidth}
\centering
\includegraphics[width=\linewidth]{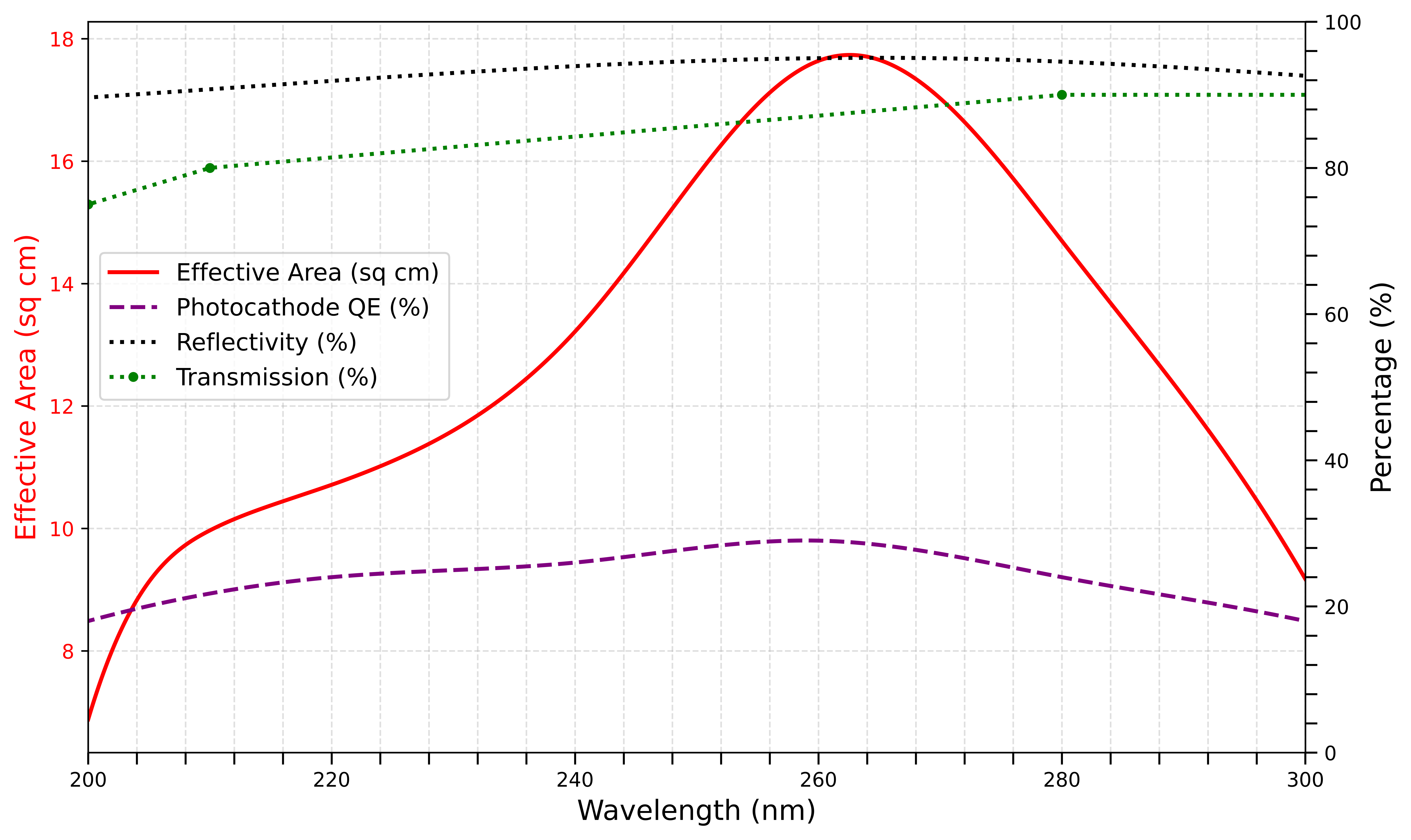}
\caption{Effective area of NUTEx (red curve) as a function of wavelength, overlaid with contributing factors: photocathode quantum efficiency (purple dashed), mirror reflectivity (black dotted), and optical transmission (green dotted).}
    \label{fig:EA_NUTEx}
    \end{minipage}
\end{figure}

While the Ritchey Chrétien design eliminates spherical aberration and coma, residual field curvature and astigmatism are present. In the baseline RC design, the geometric spot radius varies between 70 and 120 $\mu$m across the FoV. To address this, two corrective lenses were introduced and optimized in Zemax: a plano convex lens to reduce field curvature and a plano concave lens to compensate astigmatism. Their placement before the detector plane reduces the geometric spot radius to within 20 $\mu$m over the full FoV, ensuring improved image quality across the field. This performance of the optical system in terms of spot diagram is shown in Figure \ref{fig:spot diagram}.

\noindent A sensitivity tolerance analysis was conducted for NUTEx to evaluate the impact of manufacturing deviations, alignment errors, and thermal shifts on its optical performance, and to establish acceptable tolerance limits for the optical components. The analysis was performed using Zemax\textregistered, incorporating tolerance specifications provided by the component manufacturers and constraints dictated by the practical alignment strategy. The evaluation criterion required achieving 80\% encircled energy within a 20 $\mu$m radius across the field of view. The resulting tolerance limits are summarized in Table~\ref{tab:tolerance_table}.

\begin{table}[h]
\centering
\renewcommand{\arraystretch}{1.2} % Adjust row height
\setlength{\tabcolsep}{8pt} % Adjust column spacing
\caption{Tolerances for Optical Components in Manufacturing and Alignment}
\begin{tabular}{|l|l|l|c|}
        \hline
        \textbf{Tolerance Term} & \textbf{Sub Tolerance Term} & \textbf{Object} & \textbf{Tolerance} \\
        \hline
        \multirow{10}{*}{Manufacturing} 
        & \multirow{2}{*}{Radius of Curvature (\%)} & Mirrors & 1 \\
        & & Lens & 0.1 \\
        \cline{2-4}
        & \multirow{2}{*}{Thickness ($\mu$m)} & Mirrors & $\pm$100 \\
        & & Lens & $\pm$50 \\
        \cline{2-4}
        & \multirow{2}{*}{Decenter in X \& Y ($\mu$m)} & Mirrors & $\pm$50 \\
        & & Lens & $\pm$50 \\
        \cline{2-4}
        & \multirow{2}{*}{Tilt in X \& Y (arcsec)} & Mirrors & 60 \\
        & & Lens & 60 \\
        \cline{2-4}
        & \multirow{2}{*}{Surface Accuracy ($\mu$m)} & Mirrors & 0.05 \\
        & & Lens & 0.05 \\
        \hline
        \multirow{4}{*}{Alignment} 
        & \multirow{2}{*}{Decenter in X \& Y ($\mu$m)} & Mirrors & $\pm$50 \\
        & & Lens & $\pm$50 \\
        \cline{2-4}
        & \multirow{2}{*}{Tilt in X \& Y (arcsec)} & Mirrors & 60 \\
        & & Lens & 60 \\
        \hline
    \end{tabular}
    \label{tab:tolerance_table}
\end{table}

Effective area, expressed in cm\textsuperscript{2}, quantifies the total system response by representing how efficiently the telescope transmits incident light to the detector. The effective area for NUTEx can be calculated from the following equation,
\begin{equation}
A_{\text{Eff}} = A_{\text{geo}} \times R_{\text{P}}(\lambda) \times R_{\text{S}}(\lambda) \times T_{\text{L1}}(\lambda) \times T_{\text{L2}}(\lambda) \times \text{QE}_{\text{MCP}}(\lambda)\,.
\end{equation}
$A_{\text{geo}}$ is the physical collecting area, accounting for geometric factors such as the primary aperture diameter, central obscuration, and secondary mirror spider. $R_{\text{P}}$ and $R_{\text{S}}$ are the reflectivities of the primary and secondary mirrors, while $T_{\text{L1}}$ and $T_{\text{L2}}$ denote the transmission efficiencies of the two lenses. $\text{QE}_{\text{MCP}}(\lambda)$ is the wavelength-dependent quantum efficiency of the MCP detector, based on Photek specifications. Figure~\ref{fig:EA_NUTEx} shows the effective area variation of NUTEx across the 200--300~nm range, peaking at approximately 18~cm\textsuperscript{2} near 260~nm.

\begin{figure}
    \centering
\includegraphics[width=0.65\linewidth]{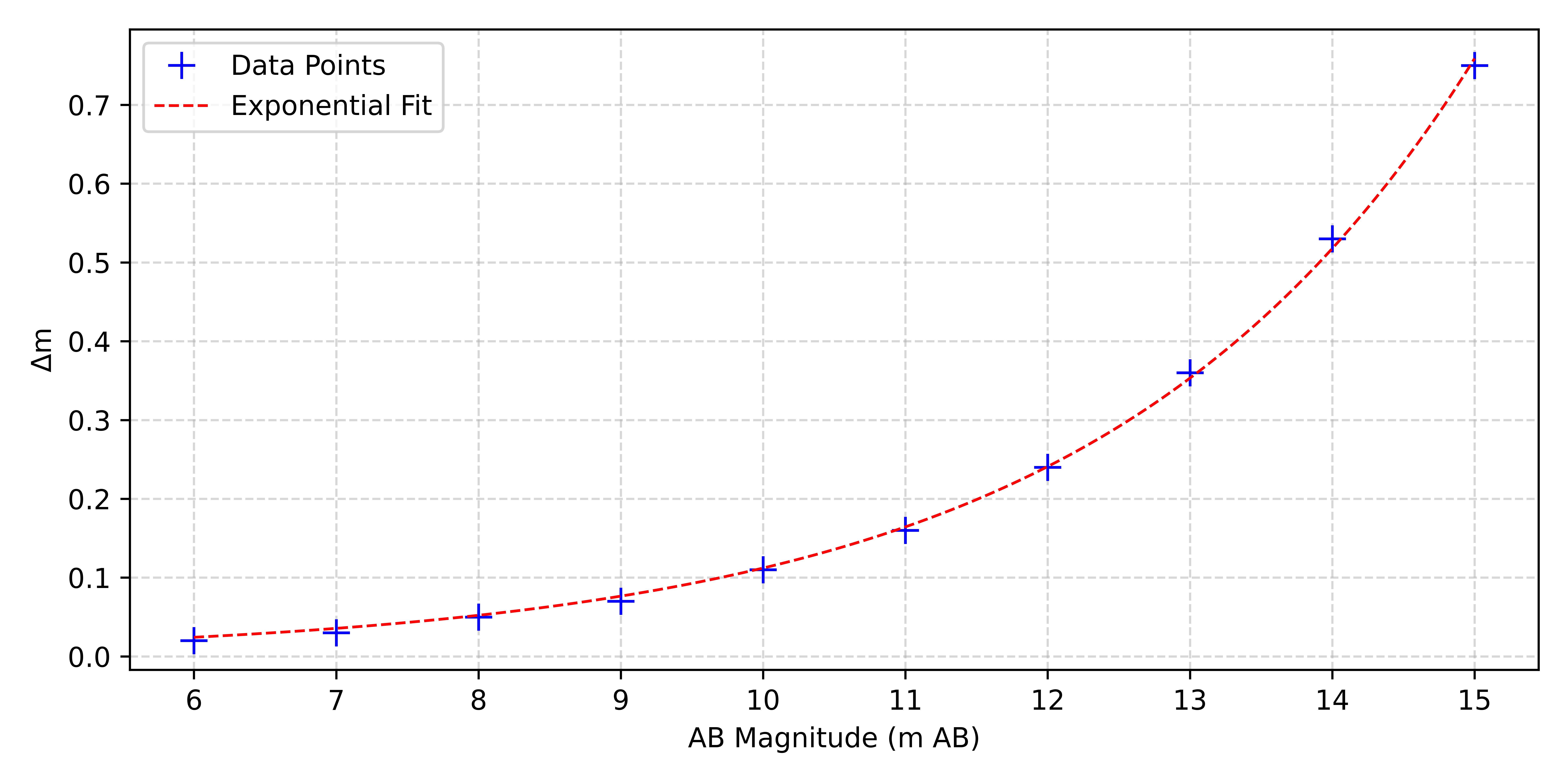}
    \caption{Photometric sensitivity of NUTEx: Minimum detectable brightness variation across AB magnitudes for a 10-second exposure at SNR = 5.}
    \label{fig:photometric accuracy}
\end{figure}

\begin{figure}
    \centering
\includegraphics[width=0.65\linewidth]{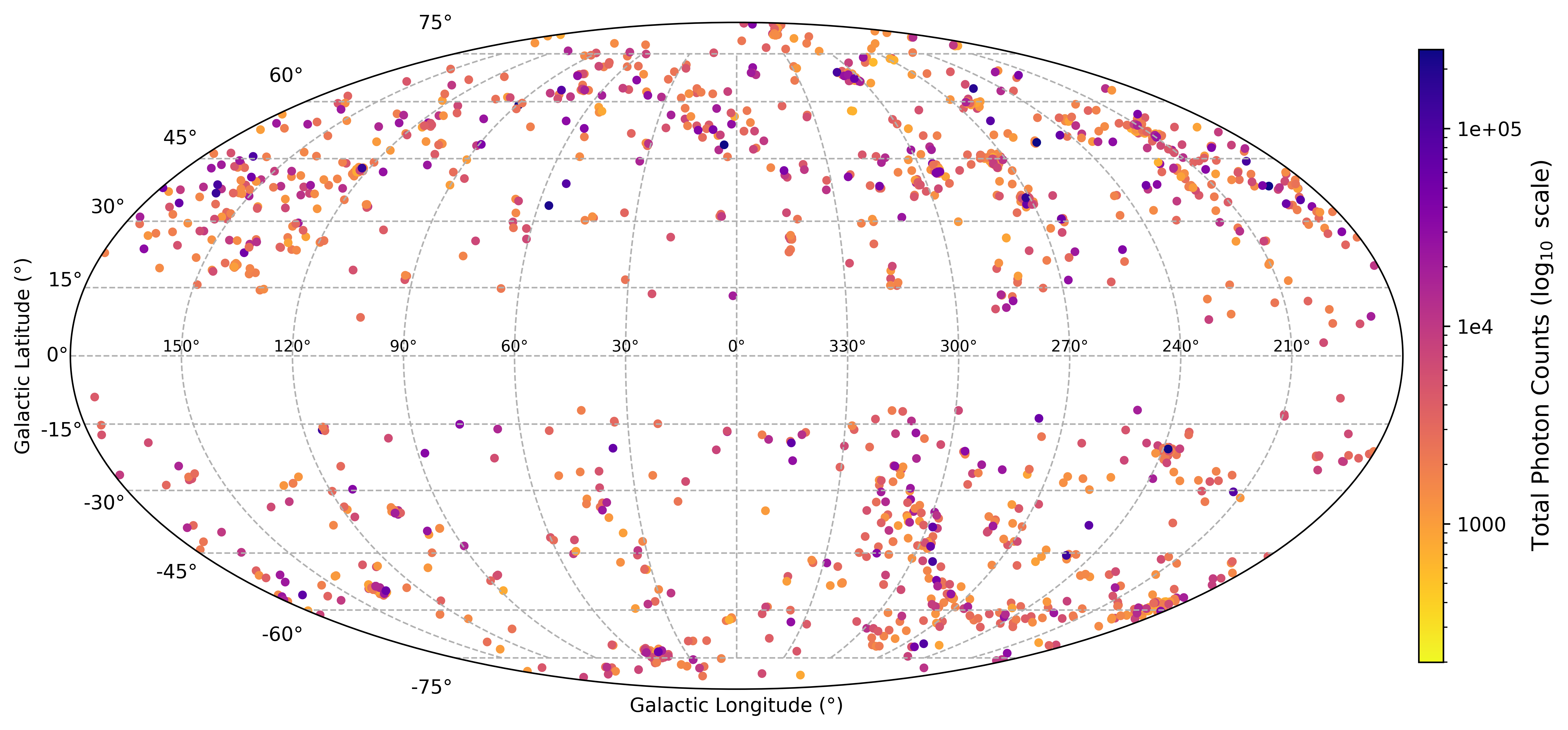}
    \caption{Simulated NUTEx photon count response for GALEX-detected UV variable sources from the GFCAT, assuming a 1K seconds exposure and the estimated effective area of NUTEx, plotted in Galactic coordinates.}
    \label{fig:galex_nuv_simulated}
\end{figure}

The photocathode’s quantum efficiency (QE) is expressed as the proportion of photoelectrons generated ($N_{\text{pe}}$) to the incoming photons ($N_{\text{photons}}$) striking its surface:
\begin{equation}
    \text{QE} = \frac{N_{\text{pe}}}{N_{\text{photons}}}\,.
    \label{eq:qe_def}
\end{equation}
The typical maximum photon gain for a specific input wavelength and phosphor screen is given by:
\begin{equation}
    G_{\gamma}(\lambda) = G_{\gamma}(\lambda_{\text{max}}) \times \left( \frac{\text{QE}(\lambda)}{\text{QE}(\lambda_{\text{max}})} \right) \times \eta_{\text{phosphor}}\,.
    \label{eq:gain}
\end{equation}
Considering $\lambda = 260$~nm, where the Photek MCP response peaks, and using $G_{\gamma}(\lambda_{\text{max}}) = 10^4$, $\text{QE}(\lambda_{\text{max}}) = 0.3$, and $\eta_{\text{phosphor}} = 1$, the maximum photon gain remains $10^4$. This implies that to maintain this gain, the number of photon events ($N_{\text{photons}}$ from Eq.~\ref{eq:qe_def}) must be limited to about 3000. The shortest exposure time of 0.033~seconds, set by the NUTEx readout unit, leads the MCP to saturate when observing stars brighter than AB magnitude 3.6, which defines the instrument’s bright limit. Using these values, we estimate the photometric accuracy for detecting transient brightness variations. For example, a 10\% variability in an AB magnitude 9.5 star can be detected with an SNR of 5 within a 10-second exposure. Figure~\ref{fig:photometric accuracy} shows the required brightness variation ($\Delta m$) across different magnitudes to achieve SNR~5 in 10-second exposure. For example, a 10\% variability in an AB magnitude 9.5 star can be detected with an SNR of 5 within a 10-second exposure. More generally, the estimated limiting sensitivity (SNR = 5) is about 9 AB magnitude for a 10-second exposure, 15 AB magnitude for 100 seconds, and 21 AB magnitude for 1200 seconds. These benchmarks indicate the range of transient events that NUTEx will be capable of detecting. To place NUTEx’s sensitivity in context, we compare with the GALEX Flare Catalog (\citet{million_gfcat}), which contains 1426 variable sources comprising flaring stars, binary systems exhibiting eclipses, pulsating stars, and active galactic nuclei (AGNs). Using reported GALEX fluxes, we simulated the corresponding photon counts as they would be detected by NUTEx. The resulting photon count distribution is shown in Figure~\ref{fig:galex_nuv_simulated}. This comparison shows that NUTEx will be capable of detecting a large fraction of GALEX-identified events, demonstrating its ability to contribute new detections of similar transient phenomena.

\subsection{Observation strategy}
NUTEx will operate as an all-sky survey instrument with public data access, rather than following proposal-driven operations typical of large observatories. Data will be released without a proprietary lock period, providing open access for the community and positioning NUTEx as a community-serving facility. A low-Earth orbit (LEO) is sufficient for the scientific goals, allowing coverage of most of the sky within approximately two years; the specific orbital parameters are not critical for the mission objectives. Pointing constraints include maintaining roughly 90° Sun avoidance, while thermal control is ensured through multilayer insulation and design validation on the engineering model over the expected temperature range. Radiation tolerance of onboard electronics, including Raspberry Pis and STM32 controllers, has been considered based on existing space heritage and the recent Starberry-Sense launch \citep{starberry2}. Observations will be carried out primarily during orbital night.

The observing strategy is designed to fulfill the primary science goals of surveying the sky for UV transients, monitoring bright regions, and conducting time-domain studies. The survey mode, combining continuous sky scanning with targeted monitoring of selected regions, allows systematic coverage to build statistically meaningful datasets for fast-evolving events such as supernova shock breakouts and stellar flares. A representative daily plan could include monitoring about five regions for one hour each, with the remainder of the orbit used for scanning. Rapid follow-up of transient alerts, such as gravitational wave triggers, for example, the GW170817 event, can be implemented as pointed observations; however, survey operations remain the core strategy, ensuring broad coverage while enabling timely response to high-priority events.

Several structural configurations were evaluated for NUTEx, including tubular, breadboard style, and modular cage designs, based on criteria such as alignment accuracy, assembly simplicity, and mass constraints. The modular cage was selected for its structural efficiency, ease of integration, and suitability for launch. Its implementation and analysis are discussed below.

\section{Mechanical design and structural analysis}

\begin{figure}
    \centering
\includegraphics[width=0.75\linewidth]{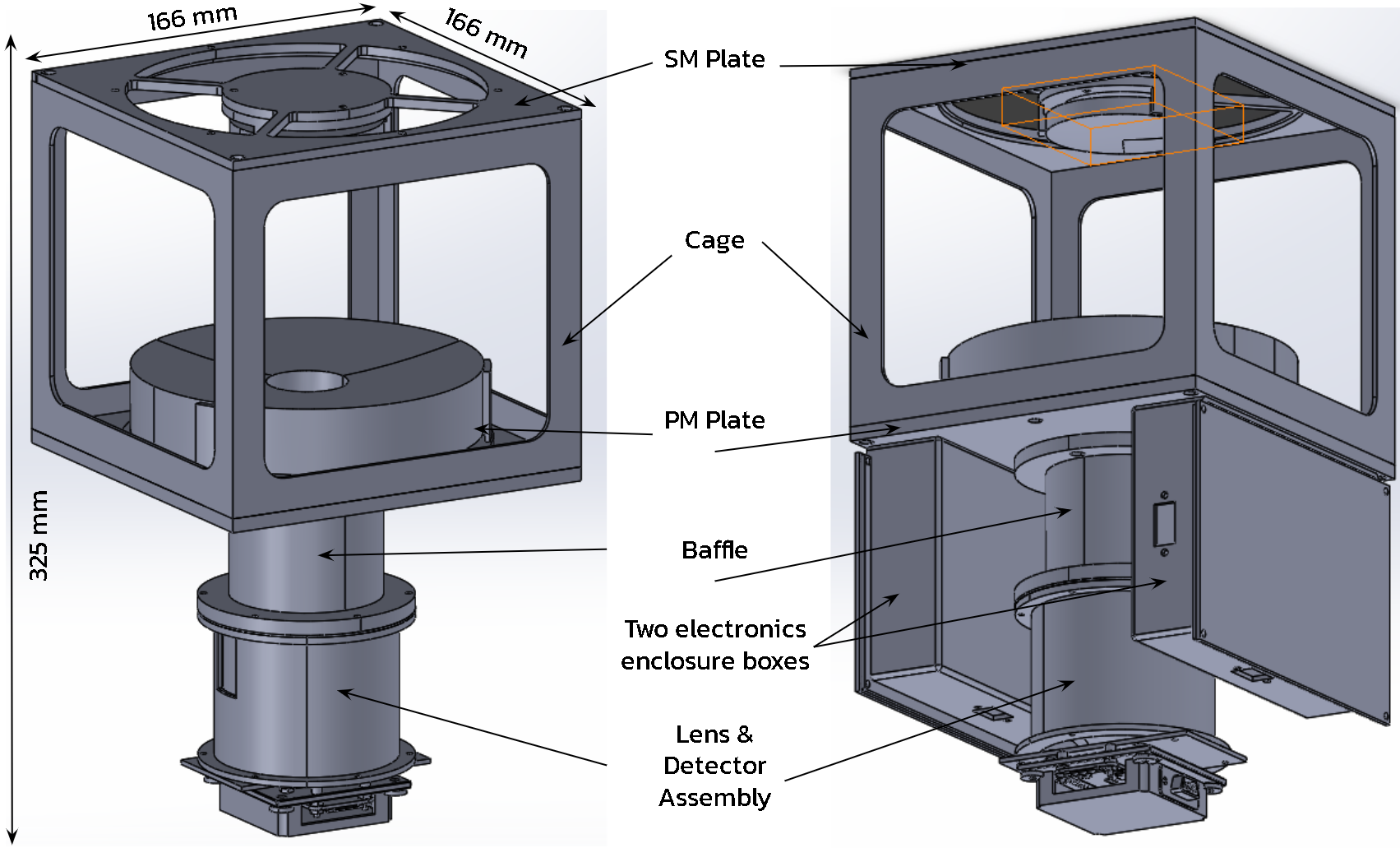}
    \caption{Fully assembled cage-based NUTEx model.}
    \label{fig:assembled model}
\end{figure}

\noindent The NUTEx payload uses a modular cage structure that reduces mass while maintaining mechanical integrity, an important factor for ensuring launch viability on low-cost missions. Although launch costs to low Earth orbit declined by about 5.5\% annually between 2000 and 2020 \citep{launch_cost}, current rates still range from \$5,000 to \$50,000 per kilogram \citep{launch_cost_2,launch_cost_3}. Drawing from prior experience with cylindrical telescope housings, we selected an open-frame cage design (see Figure~\ref{fig:assembled model}) over traditional enclosed tubes that restrict access during alignment. This configuration offers multi-sided access for integration and alignment, better accommodating clean-room procedures while simplifying mechanical layout.

The primary and secondary mirrors are bonded to mirror cells mounted on support plates at each end of the cage, with a spider-type secondary plate to reduce optical obstruction. An internal baffle supports the lens and detector assemblies while blocking stray light, eliminating the need for additional mounts. Integration involves mounting the mirrors and securing the baffle, with fine alignment adjustments made using precision shims. The onboard computer (OBC) and high-voltage power supply (HVPS) are enclosed in compact housings attached to the primary mirror plate, efficiently utilizing space. The assembled payload measures approximately 325~mm~$\times$~166~mm~$\times$~166~mm and weighs 4.326~kg, including all optical, mechanical, and electronic components. The mass distribution is 29\% optics, 54\% mechanics, and 17\% electronics. Structural parts are machined from aluminum alloy Al-6061-T6, which offers a yield strength of $2.76 \times 10^8$~Pa and a coefficient of thermal expansion of $23.6\,(\mu\text{m}/\text{m})/\text{°C})$. 

\begin{table}[h]
\centering
    \renewcommand{\arraystretch}{1.2} % Adjust row height
    \setlength{\tabcolsep}{12pt} % Adjust column spacing
\caption{Mass Budget of NUTEx Components}
\begin{tabular}{|l|c|l|c|}
        \hline
        \textbf{Component} & \textbf{Mass (Kg)} & \textbf{Component} & \textbf{Mass (Kg)} \\
        \hline
        Primary Mirror (PM) & 1.035 & Secondary Mirror (SM) & 0.075 \\
        \hline
        Lenses & 0.050 & Detector Assembly & 0.825 \\
        \hline
        PM Plate & 0.396 & SM Plate (spider) & 0.153 \\
        \hline
        Internal Baffle & 0.192 & Cage & 0.263 \\
        \hline
        Electronics Box & 0.356 & Electronics & 0.200 \\
        \hline
        Lens Housing & 0.100 & PM Cell & 0.168 \\
        \hline
        SM Cell & 0.063 & Fasteners & 0.150 \\
        \hline
        \multicolumn{4}{|c|}{\textbf{Total Mass: 4.326 Kg}} \\
        \hline
\end{tabular}
\label{tab:mass_budget_standard}
\end{table}

\subsection{Finite element analysis (FEA)}

\noindent Performing structural analyses on any CubeSat model is crucial to ensure its reliability and survivability during launch and operation. Simulation tests, such as finite element analysis (FEA) that include static analysis, thermal analysis and modal analysis help identify potential structural weaknesses and optimize the design before manufacturing, reducing cost and preventing failures \citep{Joice, Li:14}.

\subsubsection{Frequency response (modal) analysis}

\begin{figure}
    \centering
\includegraphics[width=0.75\linewidth]{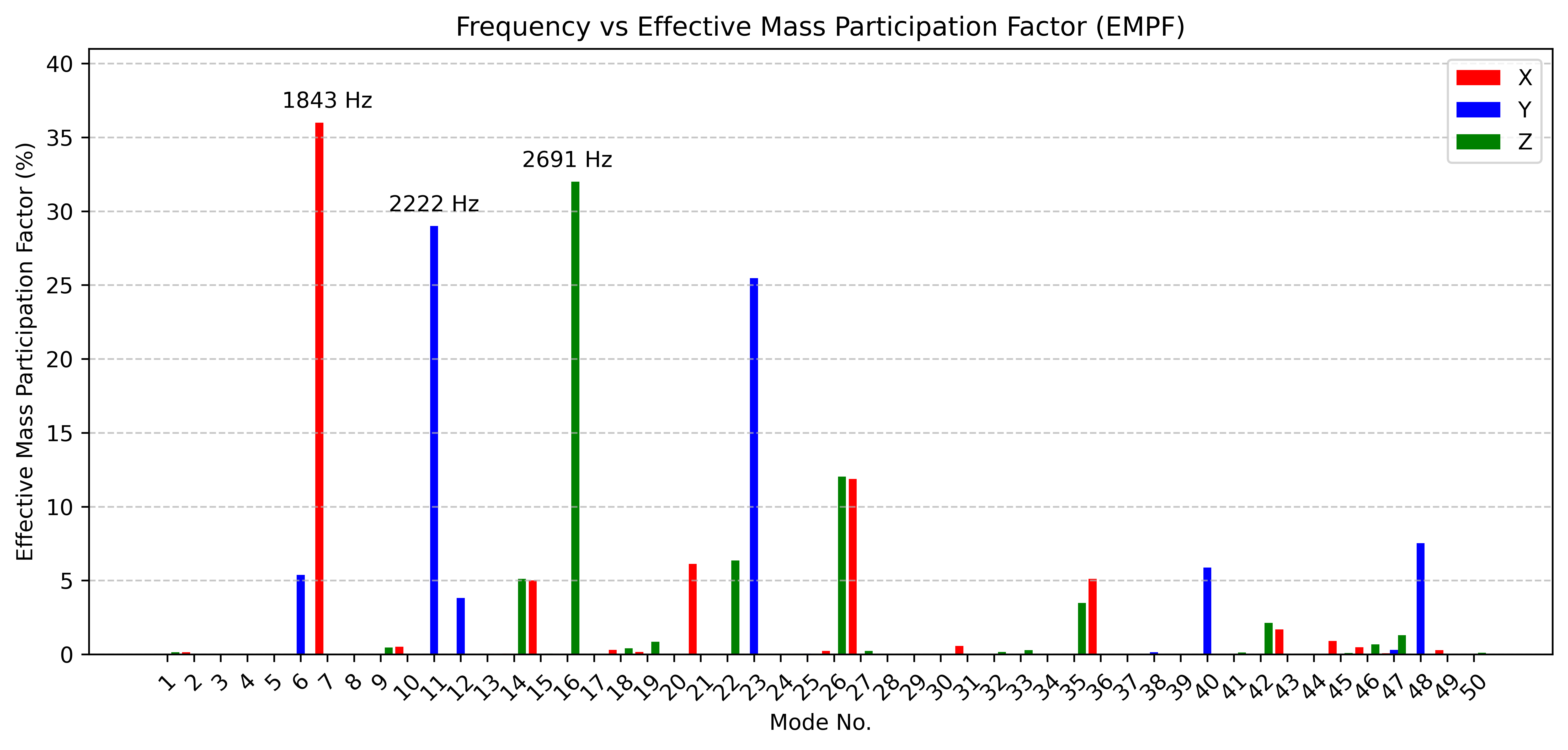}
    \caption{EMPF plot of the cage based NUTEx model}
    \label{cage empf plot}
\end{figure}

\noindent The dynamic response characteristics of the payload have been tailored to avoid resonance with the excitation profiles of the Polar Satellite Launch Vehicle (PSLV). As specified in the launcher’s guidelines, the system must exhibit primary resonant modes above 35~Hz in the longitudinal direction and 20~Hz in the lateral direction \citep{5966885, Raviprasad}. Additionally, during launches, significant harmonic frequencies occur below 100 Hz. To prevent resonance, the payload’s natural frequencies should ideally exceed 100 Hz \citep{Raviprasad}. To verify mechanical stability, a modal analysis was performed on the NUTEx structural model using mechanical design software SolidWorks, including frequency versus Effective Mass Participation Factor (EMPF) and Cumulative Mass Participation Factor (CMPF) assessments. The first ten modal frequencies are listed in Table~\ref{tab:cage modal frequencies}, with the first mode at approximately 500~Hz, well above the 100~Hz critical threshold, confirming structural adequacy. EMPF results, shown in Figure~\ref{cage empf plot}, reveal three dominant modes: Mode~7 in the X-direction at 1843~Hz (36\% participation), Mode~11 in the Y-direction at 2222~Hz (29\%), and Mode~16 in the Z-direction at 2691~Hz (32\%). The CMPF analysis up to the 50th mode shows dynamic participation of about 80\% of the total mass, supporting the robustness of the cage-based design.

\begin{table}[h!]
\centering
\caption{Modal analysis results}
\begin{tabular}{|c|c|c|c|}
        \hline
        \textbf{Mode} & \textbf{Frequency (Hz)} & \textbf{Mode} & \textbf{Frequency (Hz)} \\
        \hline
        1  & 502.79  & 6  & 1,494.6  \\
        \hline
        2  & 511.87  & 7  & 1,843.3  \\
        \hline
        3  & 621.40  & 8  & 1,859.4  \\
        \hline
        4  & 761.23  & 9  & 2,060.4  \\
        \hline
        5  & 911.11  & 10 & 2,151.0  \\
        \hline
\end{tabular}    
\label{tab:cage modal frequencies}
\end{table}

\subsubsection{Static response analysis}

\begin{figure}
    \centering
\includegraphics[width=0.65\linewidth]{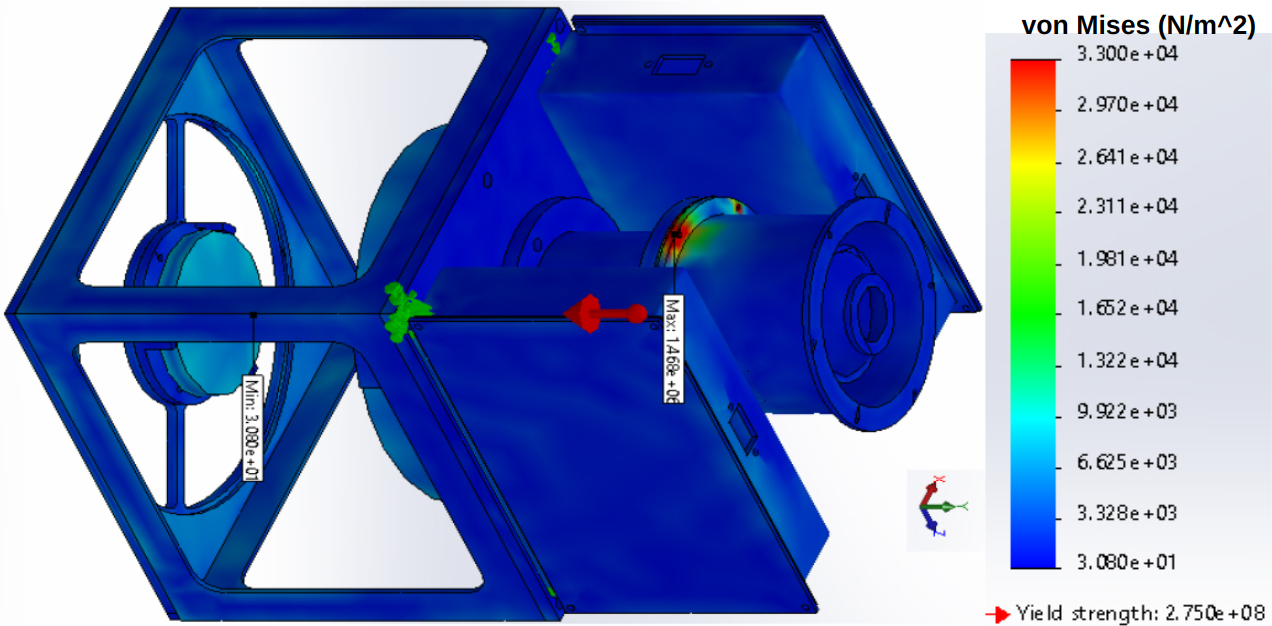}
    \caption{Cage model static analysis.}
    \label{cage-quasi-static}
\end{figure}

\noindent Static structural analysis was performed to evaluate the response of the model under quasi-static loads representative of launch conditions. A static load of 25g was applied individually along the X, Y and Z axes, and the resulting von Mises stress distributions were examined \citep{Joice}. Figure~\ref{cage-quasi-static} illustrates the stress distribution for the case when the load was applied in the Y direction, with the maximum and minimum stress locations marked. The maximum von Mises stress observed in this scenario was $1.468 \times 10^6$ N/m$^2$, as shown in Figure \ref{cage-quasi-static}. This value is significantly lower than the material yield strength, resulting in a minimum factor of safety (FoS) of approximately 200, which confirms the structural integrity of the design under quasi-static launch loads \citep{Joice}.

\section{Electronics subsystem}

\begin{figure}[htbp]
    \centering
    \begin{minipage}[t]{0.52\linewidth}
        \centering
     \includegraphics[width=\linewidth]{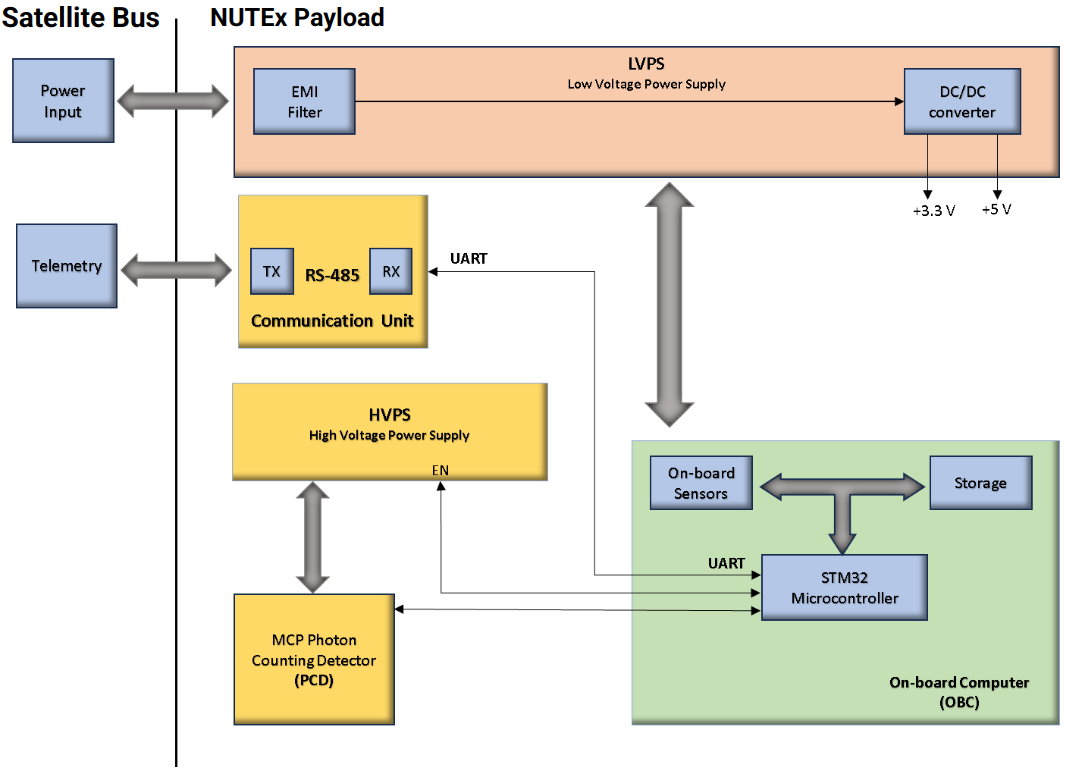}
        \caption{Block diagram of the NUTEx onboard electronics subsystem highlighting power, data handling, and communication interfaces.}
        \label{fig:elex_sub}
    \end{minipage}
    \hfill
    \begin{minipage}[t]{0.46\linewidth}
        \centering
      \includegraphics[width=\linewidth]{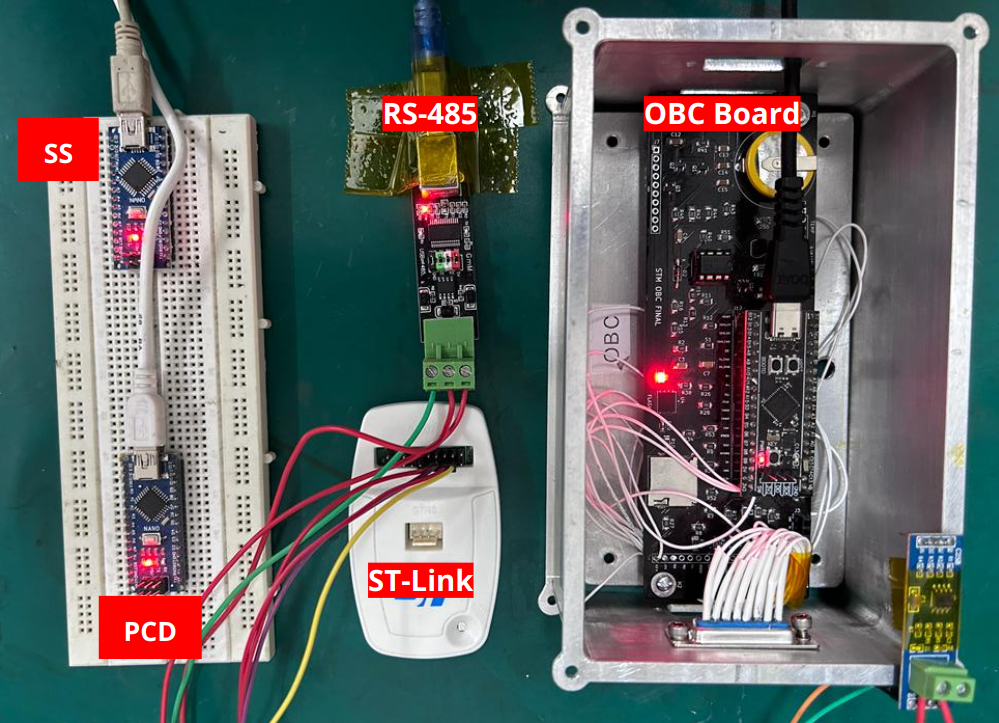}
        \caption{Test setup for validating the onboard computer and RS-485 communication interface using simulated peripheral inputs.}
        \label{fig:obc_test}
    \end{minipage}
\end{figure}

\noindent The electronics subsystem onboard a CubeSat is central to payload operations, handling data acquisition, storage, and communication with the satellite bus for command and telemetry. Recent advancements in COTS components enable the development of compact, low-cost, and power-efficient electronics tailored to specific mission needs. This has opened the way for deploying dedicated scientific instruments on resource-limited small satellite platforms. The electronics of NUTEx have been developed with these considerations in mind, as shown in Figure~\ref{fig:elex_sub}.

The payload draws power from the satellite bus, which is down-converted to required voltage levels. The OBC, based on an STM32 microcontroller, follows a compact and modular design previously demonstrated in \cite{CubeOps}, and can be adapted to NUTEx with minimal changes. Communication with the satellite bus uses an RS-485 interface, a UART-compatible protocol aligned with ISRO’s POEM platform on the PSLV \citep{starberry2}.

OBC and communication performance were validated using a laboratory setup (Figure~\ref{fig:obc_test}). Two Arduino boards emulated data from peripheral devices such as the PCD and a potential star sensor (SS). The OBC hardware and software were verified by transmitting commands and receiving data via the integrated RS-485 interface.

\section{Summary and Development schedule}

\noindent Scheduled for launch in Q2-2026, the NUTEx instrument is presented here with its optical design, science objectives, observing performance, and mechanical configuration. All optical components are manufactured, coated, and stored in a Class‑1000 cleanroom, while mechanical elements are currently in fabrication and due for completion by October-2025. Assembly, integration, and calibration will be carried out at the MGKM Laboratory, Indian Institute of Astrophysics (CREST Campus), which previously supported the UVIT payload on \textit{AstroSat} \citep{UVITmgkml}. Both engineering and flight models will be built at this facility; the engineering model will undergo vibration and thermal‑vacuum tests using the laboratory infrastructure. A full 3D-printed model of NUTEx was built and evaluated to verify the structural feasibility and integration compatibility of the design. The onboard electronics, including the OBC, power supply, and communication unit, have been developed in-house and will be further validated during system-level testing.

\begin{table}[ht]
\renewcommand{\arraystretch}{1.3} % Row height
\setlength{\tabcolsep}{8pt} % Column spacing
\centering
\caption{Key Specifications of the NUTEx Instrument}
\label{tab:nutex_specs}
\begin{tabular}{|>{\centering\arraybackslash}m{4cm}|>{\centering\arraybackslash}m{11cm}|}
\hline
\textbf{Parameter} & \textbf{Specification} \\
\hline
\textbf{Mission Class} & CubeSat-based NUV imaging payload \\
\hline
\textbf{Wavelength Range} & 200–300 nm \\
\hline
\textbf{Aperture Diameter} & 146 mm \\
\hline
\textbf{Detector Type} & 40 mm photon-counting MCP detector \\
\hline
\textbf{Field of View (FoV)} & 4° circular \\
\hline
\textbf{Spatial Resolution} & 13 arcseconds \\
\hline
\textbf{Effective Area} & Peak: 18 cm\textsuperscript{2} @260 nm; Mean: 15 cm\textsuperscript{2} across band \\
\hline
\textbf{Dimensions} & 325mm $\times$ 166mm $\times$ 166mm \\
\hline
\textbf{Mass} & 4.326 Kg \\
\hline
\textbf{Launch Readiness} & Completion by October 2025 \& Launch scheduled for Q2 2026 \\
\hline
\end{tabular}
\end{table}

NUTEx demonstrates a cost-effective approach to space-based NUV astronomy, enabling scientific observations at a fraction of the cost of larger facilities such as ULTRASAT \citep{ULTRASAT}. This affordability is made possible through a Raspberry Pi-based electronics architecture and the use of commercially available COTS components. As a pathfinder, NUTEx can inform the development of future constellations of small payloads to map the transient UV sky across the NUV and FUV bands. Its capability for extended observations of individual targets, lasting up to six months, broadens its scientific utility. The mission highlights how focused science goals can be addressed through compact platforms, providing a practical model for future instruments in this domain.

\section*{Conflict of Interest Statement}
The authors declare no competing financial or personal interests that could have influenced the work presented in this paper.

\section*{Author Contributions}

SG, RM, MS and JM contributed to writing the primary manuscript. SG and PK led the instrument's mechanical design and conducted simulation studies. MG contributed to scientific discussions. MB and SJ participated in discussions related to the electronics subsystem. JM, RM, and MS provided overall supervision and secured project funding.

\section*{Funding}

This work was supported by the Indian Institute of Astrophysics (IIA), Department of Science and Technology (DST), Government of India. 

\section*{Acknowledgments}

We express our sincere gratitude to S. Sriram and P.~U.~Kamath of the Indian Institute of Astrophysics for their invaluable suggestions and assistance. Some of the data used in this study were obtained from the Mikulski Archive for Space Telescopes (MAST). The Space Telescope Science Institute (STScI), operated by the Association of Universities for Research in Astronomy, Inc., under NASA contract NAS5-26555, manages MAST. Support for non-Hubble data is provided by the NASA Office of Space Science through grant NNX09AF08G and other funding sources.  

\section*{Data Availability Statement}
No data were generated during this research

\bibliographystyle{Frontiers-Harvard} 
\bibliography{test}
\end{document}